\documentclass{ws-procs11x85}
\usepackage{epsf}
\usepackage{amsmath}

\makeindex

\def\VEV#1{\left\langle#1\right\rangle}

\def\qq{q\bar{q}}
\def\ee{e^+e^-}
\def\MSbar{\overline{\mbox{MS}}}
\def\LQCD{\Lambda_{\mbox{\scriptsize QCD}}}
\def\as{\alpha_{\mbox{\scriptsize s}}}
\def\bas{\bar\alpha_{\mbox{\scriptsize s}}}
 
\def\al{\alpha}

\def\sig{\sigma}
\def\Sig{\Sigma}

\def\lam{\lambda}

\def\om{\omega}
\def\Om{\Omega}

\def\cV{{\cal V}}
\def\cM{{\cal M}}

\begin{document}
 
\title{Jet evolution and Monte
  Carlo\footnote{\lowercase{\uppercase{T}alk given at
      ``\uppercase{QCD} at \uppercase{C}osmic
      \uppercase{E}nergies,''\uppercase{E}rice, 29
      \uppercase{A}ugust-5 \uppercase{S}eptember, 2004.}}}
 
\author{Giuseppe Marchesini} 
 
\address{Dipartimento di Fisica, Universit\`a di Milano-Bicocca\\
INFN, Sezione di Milano, Italy}

\maketitle\abstract{In this lecture I discuss jet-shape distributions
  and describe how from jet evolution one may design Monte Carlo
  simulations which are used in the analysis of short distance
  distributions in $\ee$-annihilation, lepton-hadron and hadron-hadron
  collisions.}
 
\baselineskip=13.07pt
\section{Introduction}
A major difficulty is encountered in dealing with perturbative QCD: in
computations one uses quarks and gluons but only hadrons are observed.
Actually, there is no obvious conflict with colour confinement since
it is impossible to compute distributions with quarks or gluons in
initial or final states, they are divergent due to collinear and
infrared singularities. Even, with this general impediment, there are
ways to make predictions for hadronic distributions.
In more than thirty years since the discovery of asymptotic
freedom\cite{AF} an enormous amount of information on QCD at short
distance has been obtained by perturbative (PT) methods.
For instance, collinear (and infrared) divergences are present in DIS
structure functions (hadron in initial state) or in $\ee$
fragmentation functions (hadron in final state).  However these
divergences factorize so that one is able to compute the
evolution\cite{DGLAP} of these distributions in the hard scale and
makes predictions in terms the distribution at a given scale.

In principle, the only possibility to make absolute predictions (i.e.
without involving phenomenological inputs except for $\LQCD$) is to
compute {\it inclusive} short distance distributions such as $\ee$
total hadronic cross section in which all PT coefficients are finite
(however, as I will recall, difficulties come from non-convergence of
PT expansions).  Using inclusive observables one can reach a
``complete'' description of the emission of QCD radiation.  To give
examples of the amount of information obtained in jet-physics studies
I discuss in the following jet-shape distributions and describe how
from jet evolution one may design Monte Carlo simulations.

\section{Jet-shape observables}
In all hard processes ($\ee$-annihilation, lepton-hadron and
hadron-hadron collisions) one may introduce a large variety of
jet-shape observables. Example in $\ee$ are thrust $T$ and broadening
$B$
\begin{equation}
  \label{eq:BT}
\tau=1-T=\sum_h\frac{p_{ht}\,e^{-|\eta_h|}}{Q}\,,\qquad 
2\,B=\sum_h\frac{p_{ht}}{Q}\,,
\end{equation}
with $Q$ the hard scale (total $\ee$ center of mass energy).  The sum
is over all emitted hadrons with $p_{ht}$ transverse momentum and
$\eta_h$ rapidity with respect to the thrust axis (which maximizes
$T$). It is clear from these examples that different jet-shape
observables characterize different aspects of radiation (transverse
momentum contributes uniformly to $B$ and mostly at large angles to
$T$).  In general one finds that jet-shape observables
($V=\tau,\,B,\,C,\,D,\,K_{\rm out},\,\rho,\,y_{ij}\cdots$) are small
for most events.  In the hard process under consideration, $V\!\to\!0$
corresponds to the exclusive limit in which the minimum number of
hadrons are emitted (two for $T$ or $B$ in $\ee$) so that QCD
radiation is characterized by jets around these primary hadrons.

A quantitative description of jet radiation can be obtained by
studying the fully inclusive hadron distributions for the various
jet-shape observables $V$
\begin{equation}
  \label{eq:SigV}
  \Sig(V)=\sum_n\int\frac{d\sigma_n}{\sigma_{\rm tot}}
\Theta\left(V-\sum_n v_h\right).
\end{equation}
In the PT study of $\Sig(V)$ one replaces the sum of hadron momenta
with the one of quarks and gluons.  Is it possible to assume that the
PT distributions represent the hadron distributions? A positive answer
is suggested by the property that in general these inclusive
observables are collinear finite (given $\vec{p}_i,\,\vec{p}_j$ which
are collinear, then $V$ is the same if one replaces them with the
single momentum $\vec{p}_i+\vec{p}_i$) and infrared finite (given
$\vec{p}_i, \,\vec{p}_j$ with $|\vec{p}_i|\ll|\vec{p}_j|$ then $V$ is
the same if $\vec{p}_i$ is neglected). Therefore, assuming that
hadrons are made of collinear or soft partons, $V$ remains the same if
one replaces hadrons with partons: {\it hadron flow $\sim$ parton
  flow}. In these cases all PT coefficients of $\Sig(V)$ are finite.
One has
\begin{equation}
  \label{eq:Sig-exact}
  \Sig_{\rm PT}(V)=a(V)\,\as^{n_0}+b(V)\,\as^{n_0+1}\cdots
\end{equation}
with $\as=\as(Q)$ in $\MSbar$ scheme and $n_0$ depending on the number
of jets involved ($n_0=1$ for $2$-jet observable such as $T$ or $B$).
In the exclusive limit $V\to0$ all PT coefficient diverge due to
collinear and infrared singularities.  For $V\ne0$ these singularities
cancel but for small $V$ they leave large logarithmic terms. A
reliable PT calculation of $\Sig(V)$ at small $V$, where it is large,
requires the resummation of the most important enhanced terms. They
are organized as follows
\begin{equation}
  \label{eq:Sig-logs}
  \ln \Sig_{\rm PT}(V)=\sum_{n=1}^\infty\left\{d_n\as^n\,L^{n+1}+
s_n\as^n\,L^{n}+\cdots\right\},\qquad L=\ln V\,.
\end{equation}
In order to control the scale of the logarithms one needs to resum at
least $d_n$ (double logs, DL) and $s_n$ (single logs, SL) terms.  To
obtain a complete PT prediction for all values of $V$ one has to match
in $\Sig_{\rm PT}(V)$ both resummed \eqref{eq:Sig-logs} and the exact
\eqref{eq:Sig-exact} expressions.  Although resummation procedures are
now well established, to reach the required SL accuracy one needs very
complex tools (Mellin or/and Fourier transforms, asymptotic
estimations of integrals, etc).  Recently a numerical
program\cite{CAESAR} is available that performs automate resummations
for jet-shape distributions in $\ee$, DIS and hadron-hadron
collisions.

Before describing the structure of Feynman diagrams contributing to
the enhanced terms in \eqref{eq:Sig-logs} I will recall the
non-convergence\cite{non-conv} of PT expansion and its relation to the
large distance region of confinement.

\subsection{Power corrections}
Various physical facts are at the origin of the non convergence of PT
expansions. In general they imply the presence of power corrections to
PT results. The fact which is phenomenologically most
important\cite{power} is that the running coupling is involved at any
scale smaller than $Q$.  For example, the average value of $V$ is
given by
\begin{equation}
  \label{eq:V-average}
  \VEV{V}=\int_0^Q\frac{dk_t}{k_t}\as(k_t)\cdot \cV(k_t/Q)\,,
\end{equation}
where the virtual momentum $k_t$ in the Feynman diagrams runs into the
large distance region (although the observable is at short distance).
Since the observable is collinear and infrared finite, one has
$k_t^{-1}\cV(k_t/Q)\sim Q^{-1}$ for $k_t\to0$. Here however the
coupling enters the confinement region.  Mathematically this PT
difficulty is reflected into the fact that, although all PT
coefficients in $\as(Q)$ are finite, the expansion is non-convergent
(renormalon singularity).  To make a quantitative prediction one has
to deal with the large distance region for the running coupling. There
are various prescriptions for this.

One prescription\cite{DMW} consists in introducing a non-perturbative
(NP) parameter given by the integral of the running coupling in the
large distance region and expressing \eqref{eq:V-average} as
\begin{equation}
  \label{eq:V-average1}
  \VEV{V}=\VEV{V}^{N}_{PT}+\frac{\mu_I}{Q}\left\{C_V\,\al_0(\mu_I)+
\sum_{n=1}^N A_V^n\,\as^n\right\},\qquad
\al_0(\mu_I)=\int_0^{\mu_I}\frac{dk_t}{\mu_I}\as(\mu_I)\,,
\end{equation}
with $\mu_I$ a short distance scale ($\mu_I$-independence is ensured
by renormalization group) and $C_V$ a known constant depending on the
observable. Renormalons in the $N$-order PT expressions
$\VEV{V}^{N}_{PT}$ are canceled by the sum over the known coefficients
$A_V^n$. The contribution from the NP parameter $\al_0(\mu_I)$ is
suppressed by inverse powers in the hard scale. This power correction
is detectable\cite{power} even at LEP energies.
Similar NP contributions are found in the PT study of the distribution
$\Sig(V)$ and also here one needs to introduce the NP parameter
$\al_0(\mu_I)$. The effect here is in general a power correction
``shift'' in the argument of the PT result $\Sig_{\rm PT}(V)$.
A consequence of this prescription is that the same NP parameter
$\al_0(\mu_I)$ enters all jet-shape observables and then one can study
is phenomenological consistence. In general one finds\cite{power} that
for the various quantities the fitted values of $\al_0(\mu_I)$ varies
within about $20\%$.

Higher power corrections are present and maybe important. Here one
needs more general prescriptions\cite{KS/GR} with introduction of
shape functions to modulate large distance contributions. These
prescriptions allows one to describe the distributions at low values
of $V$.

\subsection{Structure of PT contributions}
To understand the features of QCD radiation one needs to consider how
PT resummation is obtained via factorization of (universal) collinear
and infrared singularities.  As pointed out by Dasgupta and
Salam\cite{DS}, the situation is different for global and non-global
jet-shape observables.

\vskip 0.3cm \noindent {\bf Global jet-shape observables.}  Here the
full phase space of emitted hadrons is considered. Examples in $\ee$
are $T$ and $B$ in \eqref{eq:BT}.  In these cases the DL and SL
contributions in \eqref{eq:Sig-logs} are due to gluon bremsstrahlung
emission off the primary quark-antiquark.  These collinear and/or
infrared enhanced contributions factorize and are resummed by {\it
  linear} evolution equations leading to Sudakov form factors.
Therefore, after factorization of collinear and infrared singularities
(including soft gluon coherence) QCD radiations appears as produced by
``independent'' gluon emission.  Gluon branching (into two gluons or
quark-antiquark pair) enters only in reconstructing the running
coupling as function of transverse momentum.

The fact that here the branching component does not contribute (within
SL accuracy) can be understood as a result of real-virtual
cancellations of singularities. Indeed, in the collinear limit, the
transverse momentum of an emitted gluon is equal to the sum of
transverse momenta of its decaying products. Therefore, if one
measures the total emitted transverse momentum, as in broadening for
instance, it is enough to consider the contributions of bremsstrahlung
gluons (independent emission similar to QED). Further branching does
not contribute due to unitarity (real-virtual cancellation).

\vskip 0.3cm \noindent {\bf Non-global jet-shape observables.}  Here
only a part of the phase space of emitted hadrons is considered.  The
best known example is the Sterman-Weinberg distribution\cite{SW} of energy
recorded inside a cone around a jet. A simpler example in $\ee$ is the
distribution in energy recorded {\it outside} a cone around the thrust

\vskip 0.3cm

\begin{minipage}{.4\textwidth}
\epsfig{file=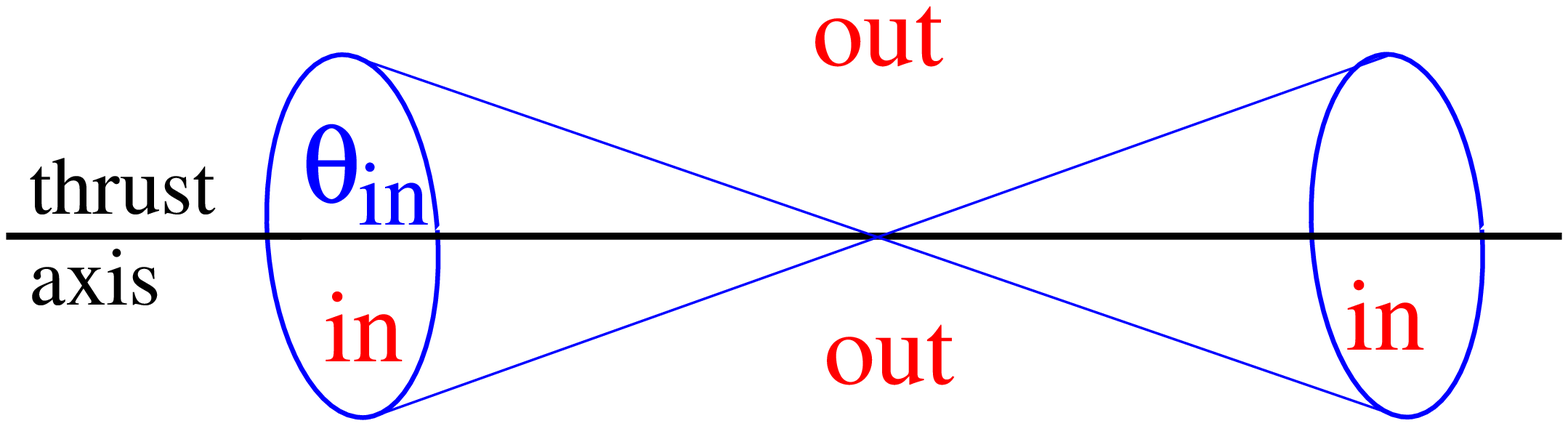,width=0.8\textwidth}
\end{minipage}
\begin{minipage}{.4\textwidth}
\begin{equation*}
  \label{eq:Eout}
\Sig(E_{\rm out})=\sum_n\int\frac{d\sig_n}{\sig_{\rm tot}}\,
\Theta\left(E_{\rm out}-\sum_{\rm out} k_{ti}\right).
\end{equation*}
\end{minipage}

\vskip 0.3cm 
\noindent
Since the jet region is excluded, there are no collinear singularities
to SL accuracy and the resummed PT contributions come from large angle
soft emission. Here resummation is more complex but informative then
in the previous cases\cite{DS}$^,$\cite{BMS}$^,$\cite{Sterman}.  Both
bremsstrahlung and branching components contribute

\vskip 0.3cm

\begin{minipage}{.4\textwidth}
\epsfig{file=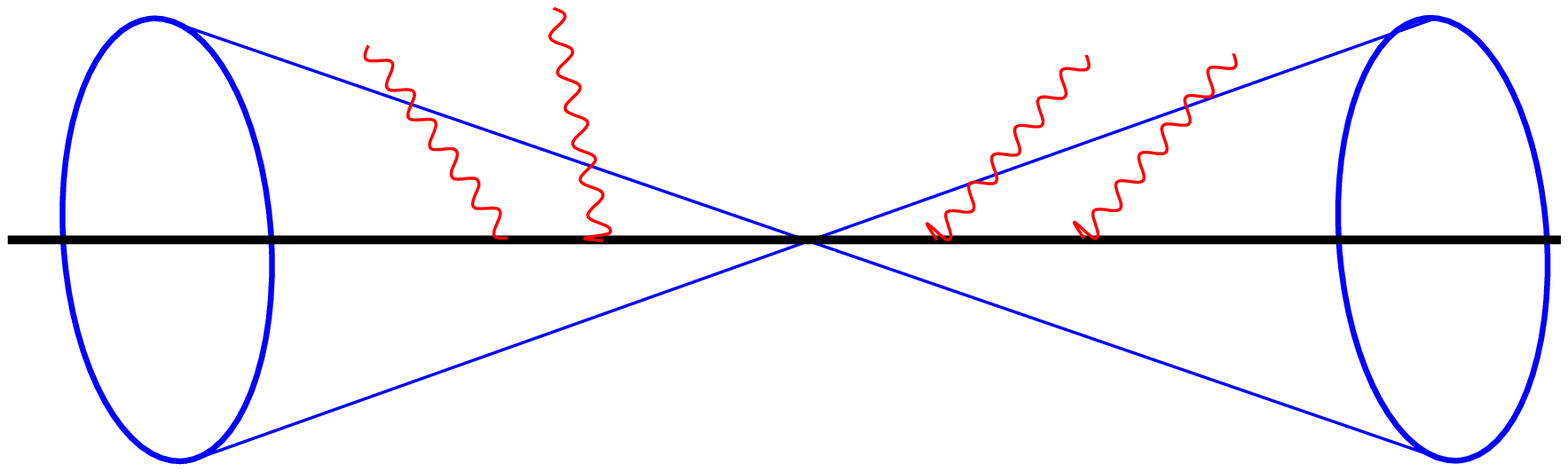,width=0.8\textwidth}
\end{minipage}
\begin{minipage}{.4\textwidth}
\epsfig{file=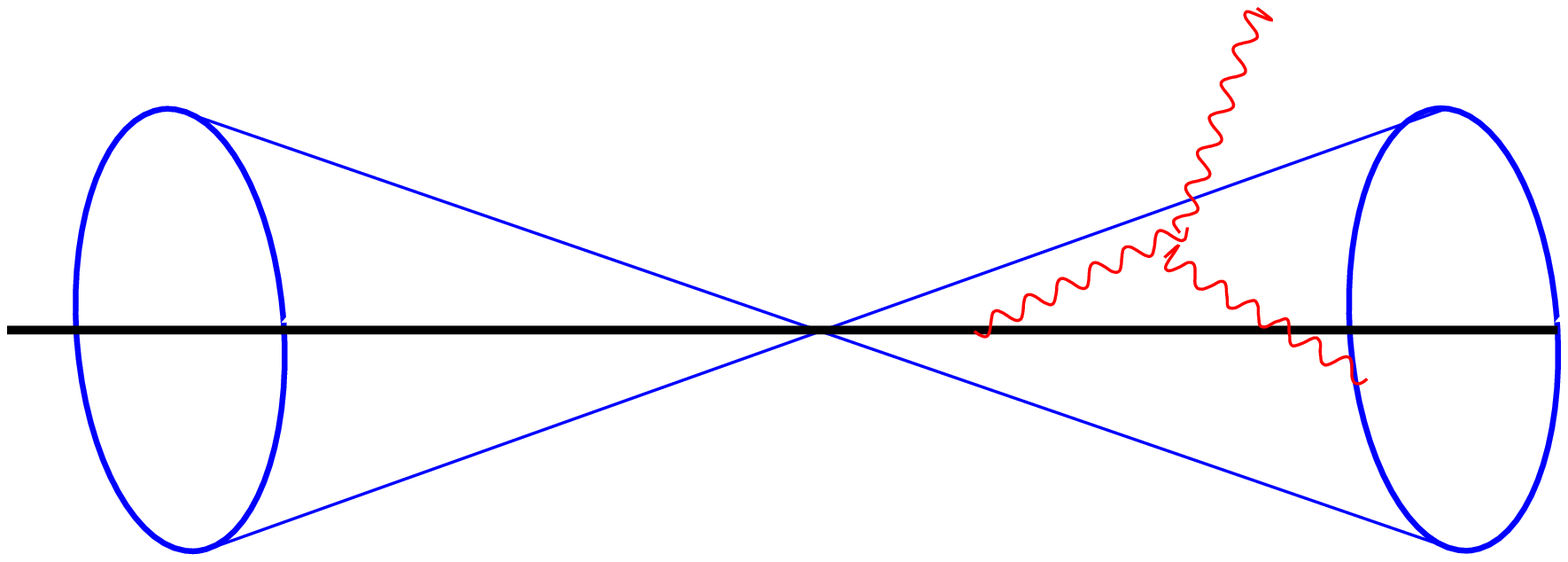,width=0.8\textwidth}
\end{minipage}
\vskip 0.3cm 
\noindent 
The bremsstrahlung component resums contributions from gluons emitted
in the recorded region outside the cone. This gives a Sudakov function
exponentially decreasing with the SL function
\begin{equation}
\label{eq:tau}
\tau=\int_{E_{\rm out}}^Q\frac{dk_t}{k_t}\,\frac{N_c\,\as(k_t)}{\pi}\,.
\end{equation}
The branching component resums contributions from gluons emitted
inside the jet region.  These gluons need to branch in order to
generate decaying products entering the recorder region.  Here
real-virtual cancellation is incomplete and virtual enhanced
contributions are dominating thus leading to a strong suppression in
the distribution which asymptotically turns out to be Gaussian in
$\tau$.

The described behaviour has been obtained by numerical\cite{DS} and
analytical\cite{BMS} studies based on the multi-soft gluon emission
distributions\cite{BCM}.  I describe this calculation which will be
used also to introduce Monte Carlo simulations for jet physics.

\section{Monte Carlo simulation}
I start by describing the derivation of the evolution
equation\cite{BMS} used to compute non-global jet-shape observables.
One introduces the {\it generating functional} for all multi-soft
gluon distributions\cite{BCM} (known only in the planar approximation
of large $N_C$) and shows that it satisfies a branching evolution
equation corresponding to a Markov process which can be numerically
implemented into a Monte Carlo simulation program. To do this one
needs to include proper cutoff for collinear and infrared
singularities.  From generated events one computes the desired
jet-shape distributions (the cutoff contributes only with power
corrections). These Monte Carlo simulations not only allows one to
compute jet-shape distributions, but provides a ``complete''
description of the hard process with soft gluons emitted in any
angular region (including large angles).  The next crucial step (apart
for including quarks and non soft contributions) consists in {\it how
  to go from parton to hadron emission}.

\subsection{Generating functional (soft and planar  limit)}
The starting point is the amplitude for the emission of $n$-soft
gluon $k_1,\cdots,k_n$ off a primary $\qq$ pair of momentum $p,\bar
p$.  It is represented as a sum of Chan-Paton factors and the 
{\it colour ordered amplitudes} coefficients
\begin{equation}
  \label{eq:BCM}
\cM_n(p\bar p q_1\cdots q_n)=\sum_{\rm perm.}
\{\lam^{a_{i_1}}\cdots \lam^{a_{i_n}}\}_{\beta\bar\beta}\>
M_n(i_1\cdots i_n)\,.
\end{equation}
From the factorization of the softest emitted gluon $q_n$
\begin{equation}
M_n(\cdots\ell\,n\,\ell'\cdots)=g_s\,M_{n-1}(\cdots\ell\ell'\cdots)\cdot 
\left(
\frac{q^{\mu}_{\ell}}{(q_{\ell}q_n)}-\frac{q^{\mu}_{\ell'}}{(q_{\ell'}q_n)}
 \right),
\end{equation}
one deduces a recurrence relation and computes all colour amplitudes
in strong energy ordering (leading order in soft limit). Summing over
colour and polarization indices, the distribution is given, to leading
$N_c$-order, by
\begin{equation}
|\cM_n|^2=\frac{1}{n!}
\prod_i\frac{N_c\as}{\om^2_i}
\sum_{\rm perm.}W_{ab}(i_1\cdots i_n),
\qquad W_{ab}(1\cdots n)\!=\!\frac{(ab)}{(aq_1)\cdots(q_nb)}\,,
\end{equation}
where $(ij)\!=\!1\!-\!\cos\theta_{ij}$ and the emission is off a
general $ab$-dipole.  Similar approximations give the multi-soft gluon
distributions in pure Yang-Mills theory. It is interesting that this
multi-soft gluon distribution coincides with the square of the exact
MHV colour amplitude discussed by Parke-Taylor\cite{MHV}.

To study arbitrary jet-shape distributions one introduces a source
function $u(q)$ for each soft gluon 
\begin{equation}
\Sigma_{ab}(E,u)=\sum_n\frac{1}{n!}\int\!
\frac{d\sig_n}{\sig_{\rm tot}}\prod_i\!{ u(q_i)}=
\sum_n\int\!\prod_i\!\left\{\!
\frac{dq_{ti}}{q_{ti}}
\frac{\,d\Om_{q_i}}{4\pi}\,{ u(q_i)}\,\bas\!\right\}
\cdot W_{ab}(1,\cdots,n)\,,
\end{equation}
with $\bas=N_c\as/\pi$ and $E=Q$.  This functional summarizes the full
information for the soft gluon emission. It involves only real
emission distribution so that it is infrared and collinear divergent.
Virtual corrections will be included later at the same accuracy in the
soft limit.

The evolution equation is obtained by using the factorization
structure of multi-soft gluon distribution
\begin{equation}
W_{ab}(1,\cdots, n)=\frac{(ab)}{(a\ell)(\ell b)}\,
W_{a\ell}(1,\cdots,\ell\!-\!1) \cdot 
W_{\ell b}(\ell\!+\!1,\cdots, n)\,,
\end{equation}
with $q_\ell$ one of the soft gluons. Taking $q_\ell$ as the hardest
(soft) gluon one obtains
\begin{equation}
\label{eq:eveq}
E\partial_E \Sigma_{ab}
=\int\frac{d\Om_q}{4\pi}\bas w_{ab}(q) 
\left[\,{ u(q)}\,\Sigma_{aq}\cdot\Sigma_{qb}-\Sigma_{ab}\,\right],
\quad w_{ab}(q)=\frac{(ab)}{(aq)(qb)}\,
\end{equation}
The negative terms in the integrand corresponds to the virtual
corrections (via Cauchy integration). They are included in the scheme
in which, for the fully inclusive case of $u(q)=1$, they completely
cancel against the real contribution ($\Sigma(E,1)\!=\!1$).  Both the
real emission branching (first term in the integrand) and the virtual
correction are collinear and infrared singular.  For inclusive
observables, (i.e.  for suitable sources $u(q)$) these singularities
cancel.

\subsection{Monte Carlo simulation, soft gluons at large angles}
The evolution equation \eqref{eq:eveq} can be formulated as a Markov
process and then numerically solved by Monte Carlo
simulations\cite{DS}.  Here the basic system is the $ab$-dipole which,
emitting a soft gluon $q$, branches into the two dipole $aq$ and
$qb$. To construct a Mante Carlo process\footnote{A Monte Carlo
program based on dipole branching similar to what is described here
has been constructed by L.  L\"onnblad\cite{LL}.} one rewrites
\eqref{eq:eveq} by splitting real and virtual corrections. To do so it
is necessary to introduce a cutoff $Q_0$ in transverse momentum (the
argument of $\as$). The Sudakov form factor
\begin{equation}
\label{eq:Sud}
\ln S_{ab}(E)\!=\!-\!\int_{Q_0}^E\!\frac{dE_q}{E_q}\int\!
\frac{d\Om_q}{4\pi}\bas w_{ab}(q)
\cdot\theta(q_{tab}\!-\!Q_0)\,,
\qquad q_{tab}^2=\frac{2E_q^2}{w_{ab}(q)}\,,
\end{equation}
is the solution of \eqref{eq:eveq} in which the real emission piece in
the integrand is neglected.  Here $q_{tab}$ is the transverse momentum
of $q$ with respect to the $ab$-dipole. Then one introduces the
probability for dipole branching: $(ab)\to(aq)\,(qb),\>\>E\!\to\!E_q$
\begin{equation}
\label{eq:branch}
d{\cal P}(E_q,\Om_q)\!=\!
\left\{\frac{dE_q}{E_q}\,\frac{S_{ab}(E)}{S_{ab}(E_q)}\right\}
\left\{\frac{d\Om_q}{4\pi}\bas w_{ab}(q)\right\}
\cdot\theta(q_{tab}\!-\!Q_0)\,.
\end{equation}
Here the first factor selects $E_q$ and the second the direction
$\Om_q$ of the emitted soft gluon. The branching does not take place
if the momentum $q_{tab}$ is below the cutoff. Each emitted dipole
then undergoes successive branchings with decreasing energy until no
further branching is permitted by the cutoff. Each Monte Carlo run
generates an ``event'' with soft gluon emitted above the cutoff.
Unitarity of probabilities ensures, in fully inclusive distribution
($u(q)\!=\!1$), the complete cancellation of real-virtual
contributions.  Computing jet-shape observables one has cancellations
of collinear and soft singularities with residual $\ln
V$-contributions and power corrections in $Q_0/(V\!E)$.  The accuracy
reached in the calculation of jet-shape observable is based on the
fact that here one uses (planar) multi-gluon emission distributions.
This implies that for jet-shape distributions one has included both DL
and SL of soft origin. SL terms of collinear origin are missing so
that here the $g\!\to\!gg$ splitting function
\begin{equation}
  \label{eq:gtogg}
P_{g\to gg}(z)=
N_c\left(\frac{1}{z}+\frac{1}{1\!-\!z} + z(1\!-\!z)\!-\!2\right),
\end{equation}
includes only the infrared singular pieces for $z\!\to\!0$ or $z\to1$.
Together with the non-soft pieces of the splitting function also
quark branching channels are missed here. 

An additional crucial missing element for a realistic jet-emission
simulation is the fact that no dipole momentum recoil is here taken
into account.  However this branching formulation correctly accounts
for soft emission also at large angle which are contributions missed
in the present Monte Carlo simulation.  It would be then important to
include recoil in \eqref{eq:branch}.

\subsection{Improved Monte Carlo simulation}
The most accurate available Monte Carlo
simulations\cite{MonteCarlo,TS,LL} are based on a branching
algorithms\cite{BCM}$^,$\cite{MW} which resum (for instance for
jet-shape distributions) DL and some of the most important SL terms.
In particular soft gluon coherence is included (angular ordering).
Continuous upgradings of Monte Carlo codes are underway which account
for new theoretical, phenomenological and experimental results. I list
here\footnote{Of course the following list does not really account for
  all the work done in the field.  It reflects my personal view of the
  important points.} some of the major features and recent (or future)
developments.

Full collinear singularity structures are included. Branching is in
general formulated as successive parton emission\cite{MonteCarlo,TS} so
that recoil and splitting functions for all parton branching ($g\to
gg,\,g\to \qq,\,q\to qg$) are naturally included. A major development
in the structure of branching is expected including resummation of
soft radiation at large angles. Partial results are however
available\cite{SSW}.

Parton branching discussed in this lecture involves only final state
emission. Similar branching holds for initial state radiation needed
in hadron-lepton or hadron-hadron collisions. However in the small-$x$
region (i.e. $Q^2\ll s$) there are peculiar differences.  Here soft
gluons are both emitted and exchanged partons. For soft exchanged
gluons angular ordering requires\cite{CCFM} additional virtual
corrections of non-Sudakov type (connected to gluon ``Reggeization'').
This brings one into the domain of ``small-$x$ physics'' in which an
accurate branching algorithms has still to be found. Partial results
are however available\cite{MW-JS}.

Using factorization structure of collinear and infrared singularities
the same branching structure holds for all hard processes. It is then
possible to construct a single Monte Carlo code for $\ee$
annihilation, lepton-hadron DIS and hadron-hadron (large $E_T$).
Moreover, such a factorization structure allows one to include in the
QCD Monte Carlo code also non-QCD processes (Electro-weak, beyond the
standard model, gravity,...). Such Monte Carlo codes are then very
useful instruments for quantitative study of ``new-physics''
scenarios.

Major recent developments in the Monte Carlo codes are systematic
attempts\cite{NLO-MC} to account for NLO and NNLO exact results
including heavy flavour processes.

\subsection {From partons to hadrons}
The above description of the Monte Carlo code refers to the generation
of events with emission of partons (possibly together with non-QCD
particles). As noted before, due to the presence of collinear and
infrared singularities these emission processes require a cutoff
$Q_0$.  The question is then how to go from partons to hadrons and how
this ``affects and distorts'' the QCD radiation.  Hadronization in the
Monte Carlo code is based on the property of
preconfinement\cite{AV}: after successive branchings, partons are
emitted in clusters of colour singlets of mass of order $Q_0$. It is
then natural to convert these colour singlet clusters into hadrons
without ``affecting and distorting'' the QCD radiation within a scale
of order $Q_0$. Preconfinement may be basis for the property that
{\it parton flow $\sim$ hadron flow}

The basis for preconfinement is again the collinear and infrared
structure of QCD. It can be explained as follows.
In the large $N_C$ limit, one may follow the colour line of partons in
the successive branchings.  The colour line of an emitted quark (or
quark part of a gluon) ends into the colour line of an emitted
antiquark (or antiquark part of a gluon). Due to this colour
connection, they form a colour singlet.
If two partons are colour connected, no emission takes place along the
colour line which connects them. So virtual corrections are dominating
and the distribution in the mass of the two colour connected partons
is suppressed by a Sudakov form factor. As a result the mass of the
two parton system is of order $Q_0$.

\section{Final considerations}

Since 1973 exact and resummed PT calculations produced enormous inside
on QCD radiation and its phenomenological ``evidence''. A clear sign
of this success is given by the Monte Carlo codes which summarize many
QCD results and are used for the analysis of data in all hard
processes (within known theoretical accuracy). Their phenomenological
success is also an indication that it is ``reasonable'' to assume that
{\it parton flow $\sim$ hadron flow}.

In PT studies it is possible to circumvent the difficulties of colour
confinement and hadronization via factorization properties or
``reasonable'' phenomenological assumption. However this is not
satisfactory on the theoretical point of view. There are indications
that the problem of colour confinement could be approached by a dual
formulation of QCD in terms of extended objects.
Lattice QCD calculations indicate\cite{Zak} that QCD vacuum is
populated by extended low dimensionality objects responsible for
confinements.  String theory provides in principle a basis for this
study. Here there are attempts\cite{string-PT} to develop new ideas
for PT calculations. Also Regge behaviour of scattering amplitudes is
studied\cite{Plochinski} and the language has some similarity with the
``Poneron'' pre-QCD topological expansion\cite{topological}.
However the natural questions are how NP formulations could account
for the enormous PT ``evidence'' and how they could inspire modeling
hadronization.

\section*{Acknowledgments}
I am grateful to the Ettore Majorana Centre for the kind invitation
and financial support to attend this very enjoyable workshop.  This
work was completed at KITP Santa Barbara, supported in part by the
National Science Foundation under grant PHY99-0794.


\begin{thebibliography}{99}
\bibitem{AF} D.J. Gross, Frank Wilczek, Phys.Rev. D8:3633,1973\\
  H.D. Politzer, Phys.Rev.Lett. 30:1346,1973
\bibitem{DGLAP} G.Altarelli, G. Parisi, Nucl.Phys. B126:298,1977\\
  V.N. Gribov, L.N. Lipatov, Yad.Fiz.15:781,1972,
  Sov.J.Nucl.Phys.15:438,1972\\
  Y.L. Dokshitzer, Sov.Phys.JETP 46:641,1977.,
  Zh.Eksp.Teor.Fiz.73: 1216,1977.
\bibitem{CAESAR} A. Banfi, G.P. Salam, G. Zanderighi,
  Phys.Lett. B584:298,2004.
\bibitem{non-conv} For reviews and classical references see V.I.
  Zakharov, Nucl.Phys. B385:452,1992; \\
  A.H. Mueller, in QCD 20
  Years Later, vol. 1 (World Scientific, Singapore, 1993)\\
  M. Beneke, Phys.Rept. 317:1,1999 [hep-ph 9807443]
\bibitem{power} For a review see for instance M.~Dasgupta and
  G.~P.~Salam, J.\ Phys.\ G{30} (2004) R143 [hep-ph/0312283].
\bibitem{DMW} Y.L. Dokshitzer, G. Marchesini, B.R. Webber,
  Nucl.Phys. B469:93,1996.  [hep-ph 9512336]
\bibitem{KS/GR} G.P. Korchemsky, G.Sterman, Nucl.Phys.B555:335,1999.
  [hep-ph 9902341]\\
  Einan Gardi, JHEP 0004:030,2000; E.Gardi, J.Rathsman,
  Nucl.Phys. B638:243,2002  [hep-ph 0201019]
\bibitem{DS} M. Dasgupta, G.P. Salam, Phys.Lett.B512:323,2001 [hep-ph
  0104277]; JHEP 0203:017,2002 [hep-ph 0203009] JHEP 0208:032,2002
  [hep-ph 0208073]; Acta Phys.Polon.B33:3311,2002 [hep-ph 0205161]
\bibitem{SW} G. Sterman and S. Weinberg, Phys.Rev.Lett. 39 (1977)
  1436
\bibitem{BMS}
  A. Banfi, G. Marchesini and G. Smye, JHEP 0208:006,2002 [hep-ph 0206076]\\
  Yu.Dokshitzer and G.Marchesini, JHEP 0303:040,2003 [hep-ph 0303101]
\bibitem{Sterman} C. Berger, T. Kucs and G. Sterman, Int. J. Mod.
  Phys.  A18:4159,2003 [hep-ph 0212343]; Phys.Rev. D68:014012,2003;
  D65:094031,2002 [hep-ph 0110004] [hep-ph 0303051]
\bibitem{BCM} A. Bassetto, M. Ciafaloni, G. Marchesini,
  Phys.Rept. 100:201,1983
\bibitem{MHV} S.J. Parke, T.R. Taylor, Phys.Rev.Lett. 56:2459,1986
\bibitem{LL}  L. L\"onnblad, Comput. Phys. Commun. 71 (1992) 15
\bibitem{MonteCarlo} G. Marchesini, B. R. Webber, G. Abbiendi, I. G.
  Knowles, M. H. Seymour and L. Stanco, Comput. Phys. Commun. 67
  (1992) 465; G. Corcella et G. Corcella, I.G. Knowles, G. Marchesini,
  S. Moretti, K. Odagiri, P. Richardson, M.H. Seymour,
  B.R. Webber,  JHEP 0101:010,2001     [hep-ph 0011363]
\bibitem{TS} T. Sj\"ostrand, Comput. Phys. Commun. 82 (1994) 74; T.
  Sj\"ostrand, P.  Eden, C. Friberg, L. L\"onnblad, G. Miu, S. Mrenna
  and E.  Norrbin, Comput. Phys. Commun. 135 (2001) 238
  [hep-ph/0010017]
\bibitem{MW} G. Marchesini, B.R. Webber, Nucl.Phys.B238:1,1984
\bibitem{SSW} S. Gieseke, P. Stephens, B. Webber, JHEP 0312:045,2003
  [hep-ph 0310083]
\bibitem{CCFM} M. Ciafaloni,  Nucl.Phys. B296:49,1988\\
  S. Catani, F. Fiorani, G. Marchesini,  Nucl.Phys. B336:18,1990\\
  G. Marchesini, Nucl.Phys. B445:49,1995 [hep-ph/9412327]
\bibitem{MW-JS} G. Marchesini and B.R. Webber, Nucl.Phys. B349:617,1991\\
  H.~Jung and G.~P.~Salam, Eur.\ Phys.\ J.\ C{19} (2001) 351
  [hep-ph/0012143]
\bibitem{NLO-MC} See for instance S. Frixione, B. R. Webber, JHEP
  0206:029,2002 [hep-ph 0204244] and S.~Frixione, P.~Nason and
  B.~R.~Webber, JHEP 0308:007,2003 [hep-ph/0305252]
\bibitem{AV} D. Amati and G. Veneziano, Phys.Lett. B83:87,1979
\bibitem{Zak} V.I. Zakharov, Dual string from lattice Yang-Mills
  theory. NSF-KITP-04-137 (Jan 2005) [hep-ph/0501011]
\bibitem{string-PT} E.~Witten, Commun.\ Math.\ Phys.\ {252} (2004) 189
[hep-th/0312171]
\bibitem{Plochinski} J.~Polchinski and M.~J.~Strassler, JHEP 0305:012,
  2003 [hep-th/0209211]; Phys.\ Rev.\ Lett.\ {88} (2002) 031601
  [hep-th/0109174]
\bibitem{topological} M. Ciafaloni, G. Marchesini, G. Veneziano,
  Nucl.Phys.B98:472,1975; B98:493,1975
\end{thebibliography}
\end{document}